\documentclass[twocolumn]{aastex631}
\usepackage{xfrac}
\usepackage{amsmath,graphicx,hyperref}
\usepackage{multirow}
\usepackage{soul}
\usepackage{longtable}
\submitjournal{ApJ}
\shorttitle{}
\shortauthors{}
\def\new#1{{\color{black}#1}}
\def\neww#1{{\color{black}#1}}

\begin{document}

\title{The Cosmic Ultraviolet Baryon Survey (CUBS) IX: The enriched circumgalactic and intergalactic medium around star-forming field dwarf galaxies traced by \ion{O}{6} absorption}

\correspondingauthor{}
\email{mishran@umich.edu}
\textit{}
\author[0000-0002-9141-9792]{Nishant Mishra}
\affiliation{Department of Astronomy, University of Michigan, 1085 S. University, Ann Arbor, MI 48109, USA}

\author[0000-0001-9487-8583]{Sean D. Johnson}
\affiliation{Department of Astronomy, University of Michigan, 1085 S. University, Ann Arbor, MI 48109, USA}

\author[0000-0002-8459-5413]{Gwen C. Rudie}
\affiliation{The Observatories of the Carnegie Institution for Science, 813 Santa Barbara Street, Pasadena, CA 91101, USA}

\author[0000-0001-8813-4182]{Hsiao-Wen Chen}
\affiliation{Department of Astronomy \& Astrophysics, The University of Chicago, Chicago, IL 60637, USA}

\author[0000-0002-0668-5560]{Joop Schaye}
\affiliation{Leiden Observatory, Leiden University, PO Box 9513, NL-2300 RA Leiden, The Netherlands}

\author[0000-0002-2941-646X]{Zhijie Qu}
\affiliation{Department of Astronomy \& Astrophysics, The University of Chicago, Chicago, IL 60637, USA}

\author[0000-0001-7869-2551]{Fakhri S. Zahedy}
\affiliation{Department of Physics, University of North Texas, Denton, TX 76203, USA}
\affiliation{The Observatories of the Carnegie Institution for Science, 813 Santa Barbara Street, Pasadena, CA 91101, USA}

\author[0000-0003-3244-0409]{Erin T. Boettcher}
\affiliation{Department of Astronomy, University of Maryland, College Park, MD 20742, USA}
\affiliation{X-ray Astrophysics Laboratory, NASA/GSFC, Greenbelt, MD 20771, USA}
\affiliation{Center for Research and Exploration in Space Science and Technology, NASA/GSFC, Greenbelt, MD 20771, USA}

\author[0000-0001-5804-1428]{Sebastiano Cantalupo}
\affiliation{Department of Physics, University of Milan Bicocca, Piazza della Scienza 3, I-20126 Milano, Italy}

\author[0000-0002-8739-3163]{Mandy C. Chen}
\affiliation{Department of Astronomy \& Astrophysics, The University of Chicago, Chicago, IL 60637, USA}

\author[0000-0002-4900-6628]{Claude-Andr\'e Faucher-Gigu\`ere}
\affiliation{Department of Physics \& Astronomy, Center for Interdisciplinary Exploration and Research in Astrophysics (CIERA), Northwestern University, 1800 Sherman Avenue, Evanston, IL 60201, USA}

\author[0000-0002-5612-3427]{Jenny E. Greene}
\affiliation{Department of Astrophysics, Princeton University, Princeton, NJ 08544, USA}

\author[0000-0002-0311-2812]{Jennifer I-Hsiu Li}
\affiliation{Department of Astronomy, University of Michigan, 1085 S. University, Ann Arbor, MI 48109, USA}
\affiliation{Michigan Institute for Data Science, University of Michigan, Ann Arbor, MI 48109, USA}

\author[0000-0002-2662-9363]{Zhuoqi (Will) Liu}
\affiliation{Department of Astronomy, University of Michigan, 1085 S. University, Ann Arbor, MI 48109, USA}

\author{Sebastian Lopez}
\affiliation{Departamento de Astronom\'ia, Universidad de Chile, Casilla 36-D, Santiago, Chile}

\author{Patrick Petitjean}
\affiliation{Institut d’Astrophysique de Paris 98bis Boulevard Arago, 75014, Paris, France}

\begin{abstract}
The shallow potential wells of star-forming dwarf galaxies make their surrounding circumgalactic and intergalactic medium (CGM/IGM) sensitive laboratories for studying the inflows and outflows thought to regulate galaxy evolution. We present new absorption-line measurements in quasar sightlines probing within projected distances of \new{$<300$} kpc \new{from} \neww{91} star-forming field dwarf galaxies with a median stellar mass of $\log{M_\star/\rm{M_\odot}} \approx 8.3$ at $0.077 < z < 0.73$ from the Cosmic Ultraviolet Baryon Survey (CUBS). In this redshift range, the CUBS quasar spectra cover a suite of transitions including \ion{H}{1}, low and intermediate metal ions (e.g., \ion{C}{2}, \ion{Si}{2}, \ion{C}{3}, and \ion{Si}{3}), and highly ionized \ion{O}{6}.
%The CUBS-Dwarfs sample enables such multi-phase CGM/IGM studies in dwarf samples 9$\times$ larger than past work. 
\new{This CUBS-Dwarfs survey enables constraints with samples} \neww{9}\new{$\times$ larger than past dwarf CGM/IGM studies with similar ionic coverage.}
We find that low and intermediate ionization metal absorption is rare around dwarf galaxies, consistent with previous surveys of local dwarfs.
In contrast, highly ionized \ion{O}{6} is commonly observed in sightlines that pass within the virial radius of a dwarf, and  \ion{O}{6} detection rates are non-negligible at projected distances of 1$-$2$\times$ the virial radius. Based on these measurements, we estimate that the \ion{O}{6}-bearing phase of the CGM/IGM accounts for a dominant share of the metal budget of dwarf galaxies.
The absorption kinematics suggest that a relatively modest fraction of the \ion{O}{6}-bearing gas is \new{formally} unbound. Together, these results imply that low-mass systems at $z\lesssim 1$ effectively retain a substantial fraction of their metals within the nearby CGM and IGM.
\end{abstract}

\section{Introduction}
\label{sec:intro}

The intergalactic medium (IGM) and circumgalactic medium (CGM) play an important role in galaxy evolution by acting as significant baryon and metal reservoirs \citep[for a review, see][]{Tumlinson17}. Moreover, simulations suggest that the physical state of the CGM and IGM are critical for understanding how galaxies sustain star formation and experience quenching \citep[for a review, see][]{Crain:2023}. Hydrodynamic simulations of galaxy evolution \citep[\new{e.g.,}][]{AnglesAlcazar17, Mitchell22}, together with observations of both inflows \citep[\new{e.g.,}][]{Rubin12} and outflows \citep[\new{e.g.,}][]{Rupke19}, indicate a dynamic CGM with repeated cycling of baryons between star-forming regions in galaxies and their surrounding cosmic ecosystems.
However, the physical processes driving these cycles and resulting fractions of halo gas undergoing recycling, outflow, or inflow remain unresolved \citep[for a review of physical processes in the CGM, see][]{FaucherGiguereOh23}.

The shallow potential wells of dwarf galaxies \neww{\citep[defined as $\log M_\star / {\rm M_\odot} < 9$ in the review by][]{Bullock17}} are expected to make the properties of their surrounding CGM and IGM sensitive to feedback mechanisms. Indeed, hydrodynamic simulations of dwarf galaxies with different \new{stellar} feedback recipes result in dramatically different extents of metal enrichment around dwarf halos. 
While some highly ejective supernova feedback prescriptions predict that dwarf halos will lose a substantial fraction of their metals and baryons to the IGM on Mpc scales \citep[e.g.,][]{Mina21}, less aggressive feedback recipes are predicted to produce galactic fountains around dwarf galaxies, resulting in the retention of baryons within their CGM \citep[e.g.,][]{Muratov17}. In larger cosmological simulations, these differences in feedback prescription manifest themselves in large differences in the fraction of halo gas contained within the CGM, $f_{\rm CGM}$ \new{\citep{Davies20}}. However, the observations of the CGM and IGM around dwarf galaxies needed to test these predictions are challenging due to the low luminosities of the galaxies and the expected small sizes of their halos.

The densities ($\log n_{\rm H} / \rm cm^{-3} \approx -5$ to $0$) and temperatures ($\log T / \rm K \approx 4$ to $6$) of the CGM and IGM often make observing the gas in emission with existing facilities \new{challenging and limited to the densest phases \citep[e.g.,][]{Guo23}}. Therefore, sensitive rest-frame UV absorption-line spectroscopy represents the primary method for probing this normally invisible gas. Since the installation of the Cosmic Origins Spectrograph \citep[COS; ][]{GreenCOS12} aboard the \textit{Hubble Space Telescope (HST)}, observations of the CGM at $z\lesssim 0.2$ have shown that it is multi-phase, i.e. contains complex ionization states \citep[e.g.,][]{Giroux94}. However, the majority of empirical surveys of the CGM at modest redshift have focused on massive and intermediate-mass galaxies.

The combination of the increased UV sensitivity of COS and wide-field surveys have resulted in a revolution of our understanding of the multiphase CGM of massive galaxies at $z<0.4$ from luminous red galaxies \citep[$\log M_\star/{\rm{M_\odot}} > 11$; e.g.,][]{Chen18COSLRG, Chen19COSLRG, Zahedy19COSLRG, Smailagic18, Berg19RDR, Smailagic23}, to Milky Way-mass systems \citep[$\log M_\star/{ \rm{M_\odot}} \sim 10$; e.g.,][]{Stocke13, Tumlinson13COSHalos, Mathes14, Johnson15, Borthakur15, Borthakur16, Prochaska19CASBAH, Wilde21CGM2, Tchernyshyov22}. Similar survey approaches at somewhat lower redshift have led to significant advances in our understanding of the gaseous halos around intermediate-mass galaxies of $\log M_\star/{\rm{M_\odot}} \new{\approx} 9.5$ \citep[e.g.,][]{Bordoloi14, Bordoloi18}. However, our empirical knowledge of the CGM and IGM around low-mass dwarf galaxies is more uncertain.

The majority of studies of the CGM of dwarf galaxies ($\log M_\star/{\rm{M_\odot}} \lesssim \new{9}$) have focused on the local Universe due to the difficulty in detecting faint galaxies at \new{$z>1$, with the exception of Lyman-$\alpha$ emitters (LAEs) at $z>2$ \citep[e.g.,][]{Zahedy19LAE, Muzahid21, Banerjee23, Lofthouse23}. These surveys  \citep[e.g.,][]{LiangChen14, QuBregman19, Zheng20, Zheng23}} found that low and intermediate ionization state metal detections of \ion{Si}{2}, \ion{Si}{3}, \ion{Si}{4}, \ion{C}{2}, and \ion{C}{4} are rare and represent a modest fraction of the metal budget of dwarf systems. However, low-redshift CGM observations have limited rest-frame wavelength coverage of the key ionic transitions such as the \ion{O}{6} $\lambda\lambda 1031, 1037{\rm \AA}$\ doublet, which is the most abundant metal line detected in the $z\lesssim 0.7$ Universe \citep[e.g.,][]{Danforth16}. Furthermore, such local CGM studies are unable to robustly constrain \ion{H}{1} column density due to contamination from the damping wings of Ly$\alpha$ absorption arising from the ISM of the Milky Way and lack of coverage of higher-order Lyman series lines. Therefore, to access a wider range of absorption lines that constrain additional CGM/IGM phases, we must move to intermediate redshifts of $z\gtrsim 0.1$ due to the low throughput of the {\it HST} optical system at $\lambda <1100$ \AA.

%Johnson 2017
In previous work observing low-mass galaxies at $z\gtrsim0.1$, \cite{Johnson17} examined the CGM/IGM around 18 star-forming dwarf galaxies, finding low detection rates of \ion{Si}{2}, \ion{Si}{3}, \ion{Si}{4}, \ion{C}{4}, but a $\sim 50 \%$ detection rate of \ion{O}{6} in sightlines with projected distances, $d_{\rm proj}$, less than the dwarf galaxy virial radii, $R_{\rm vir}$. Recently, deep galaxy surveys around intermediate redshift quasars with COS spectra \citep{Tchernyshyov22, Qu24CUBSVII} found similar detection rates of \ion{O}{6} around dwarf galaxies. This comparatively high detection rate of \ion{O}{6} around dwarf galaxies suggests that their CGM is dominated by high ionization state gas and represents a significant portion of the metal budget of dwarf galaxies. However, larger samples are necessary to better characterize the multi-phase CGM and IGM around low-mass systems.

%environment

When conducting observational studies to connect the properties of galaxies with absorption systems detected around them, controlling for galaxy environment is critical. Observations of the star-formation rates of dwarf galaxies indicate a strong correlation between quiescence and environment \citep[e.g.,][]{Peng10, SlaterBell14, SlaterBell15, Green24ELVESIII}. Simulations \citep[e.g.,][]{Williamson18, Emerick16, Zhu24, Roy24} suggest that this may be due to the ease of stripping dwarf galaxies of their CGM and ISM during interactions. Direct observations of the \ion{H}{1} emission in the local volume \citep[e.g.,][]{Kleiner23MeerKAT} and LMC/SMC in the local group \citep[e.g.][]{Salem15} also demonstrate the effectiveness of ram pressure in removing gas from dwarf galaxies as they interact with more massive ones. Similarly, absorption properties have been observed to be highly dependent on the galactic environment \citep[e.g.,][]{ChenMulchaey09, MulchaeyChen09, Johnson13, Johnson18, Burchett16, CUBSI, Beckett21, Huang21, Dutta21}. \new{For example, \cite{Dutta24} found an excess of \ion{H}{1} absorption for galaxies in group environments versus isolated systems at $z<0.5$.}

Therefore, developing an empirical understanding of the role of stellar feedback in shaping baryon cycles in and around dwarf galaxies requires observations of the CGM of field dwarfs where these environmental effects from interactions with massive neighbors are minimal. Recent $z \lesssim 1$ CGM/IGM surveys such as CASBAH \citep{Burchett19CASBAHI}, CUBS \citep{CUBSI}, MUSEQuBES \citep{MUSEQuBES}, and CGM$^2$ \citep{Wilde21CGM2} work to accomplish this by conducting deep and highly complete galaxy redshift surveys in the fields around quasar sightlines with COS UV spectra.

In this work, we leverage high-quality COS UV spectroscopy as well as deep and highly complete galaxy redshift surveys from the Cosmic Ultraviolet Baryon Survey (CUBS), to construct a sample of \neww{91} isolated, star-forming dwarf galaxies for which we can measure the surrounding CGM and IGM with a multi-phase suite of associated absorbers with quasar sightlines at $d_{\rm proj} < 300$ kpc. \new{Throughout the paper, we express all distances in proper (physical) units unless stated otherwise. For calculations with cosmological dependence, we use the standard $\Lambda$-cosmology with $H_0 = 70$ km\,s$^{-1}$\,Mpc$^{-1}$, $\Omega_M = 0.3$, and $\Omega_{\Lambda} = 0.7$.} The paper is organized as follows. In Section \ref{sec:metho}, we briefly review the methodology of the CUBS survey and the properties of the sub-sample of the survey consisting of isolated dwarf galaxies, hereafter referred to as CUBS-Dwarfs. We also discuss the inference of absorber properties via Voigt profile fitting. In Section \ref{sec:results}, we characterize the CGM and IGM of the dwarf galaxies, focusing on both total ion columns and covering fractions. In Section \ref{sec:discussion}, we discuss the implications of our results with respect to the metal and mass budgets of the CGM of dwarf galaxies. Finally\new{,} in Section \ref{sec:summary}, we summarize our conclusions.

\section{Methodology}
\label{sec:metho}

\subsection{Dwarf galaxy sample}
\label{sec:galsample}

The CUBS survey was designed to enable unbiased studies of the CGM and IGM at $z\lesssim 0.8$ by obtaining high-quality COS FUV spectroscopy of 15 NUV-selected quasars at $z\gtrsim0.8$. Here, we briefly review the CUBS survey design and strategy, more details of which can be found in CUBS Paper I \citep{CUBSI}. The CUBS quasars were selected in an absorption-blind manner using the GALEX near-UV bandpass (NUV; 1770-2730\rm \AA). This NUV selection was necessary to avoid biasing the survey against Lyman Limit Systems (LLS), \ion{H}{1} absorbers with $\log{N / \rm cm^2} > 17$, which are optically thick to the Lyman continuum. These systems are important probes of the cool, dense phases of the CGM \citep[e.g.,][]{Lehner18CCC1, Chen19COSLRG, Wotta19CCC2LLS, CUBSI} but also attenuate FUV flux.

FUV spectra of the 15 NUV-selected quasars were observed between 2018 and 2019, as part of the HST Cycle 25 (PID=15163; PI: Chen). We observed the quasars with the G130M and G160M gratings on COS, with integration times designed to achieve signal-to-noise (S/N) per 20 $\rm km\,s^{-1}$\ resolution element of $\approx 15$ to ensure sensitivity. The spectral range of COS ($\lambda \approx 1100\new{\rm{\AA}}-1800\new{\rm{\AA}}$) provides access to a range of atomic transitions from $0.1 \lesssim z \lesssim 0.7$ such as the Lyman-series for \ion{H}{1}, as well as multiphase metal ions including low (\ion{C}{2}, \ion{Si}{2}), intermediate (e.g., \ion{C}{3}, \ion{Si}{3}) and high (e.g., \ion{O}{6}) ions. This opens up the potential to estimate the temperature and density of CGM gas at a range of redshifts, especially at $z > 0.2$, where we have access to all the \ion{H}{1} transitions up to the Lyman limit \citep[e.g.,][]{Zahedy21CUBSII, Cooper21CUBSIV, Qu22CUBSV, Qu23CUBSVI,Qu24CUBSVII}. 

\begin{figure*}
\includegraphics[width=\textwidth]{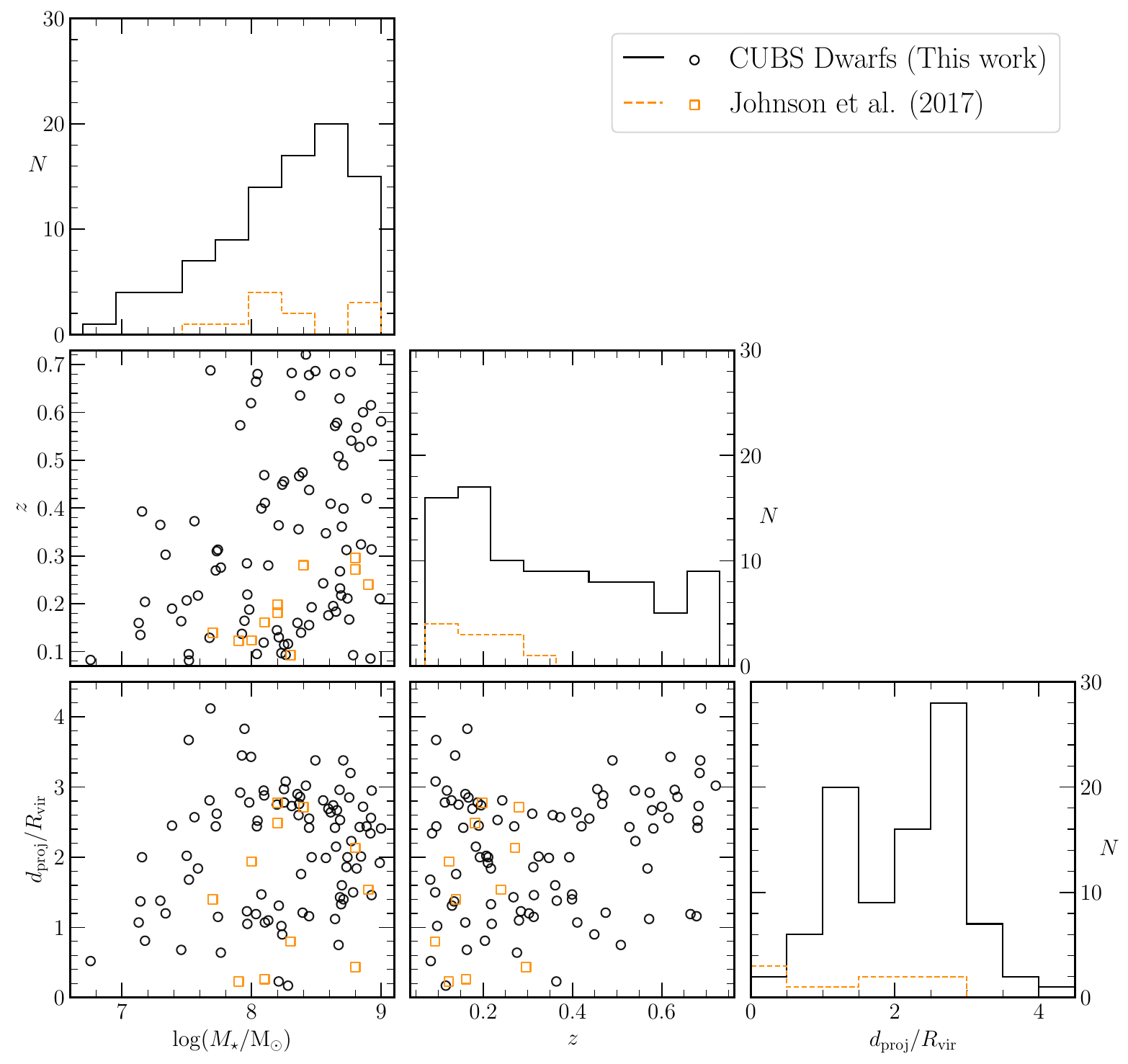}
\caption{The stellar mass ($\log{M_\star/\rm{M_\odot}}$), redshift ($z$), and virial radius normalized project distance ($d_{\rm proj}/R_{\rm vir}$) distributions of the sample plotted against each other with histograms plotted along the diagonal. The CUBS-Dwarfs sample is shown in circles and solid lines while the dwarf sample from \cite{Johnson17} is shown with \new{orange} square symbols and dotted lines for comparison. The \neww{91} CUBS-Dwarfs galaxies enable studies of the multi-phase CGM/IGM around dwarf galaxies with a 9-fold increase in sample size, including an emphasis on $d_{\rm proj} / R_{\rm vir} > 1$ and $z > 0.3$.} 
\label{fig:sample}
\end{figure*}

To connect our absorber catalog with galaxy properties, we performed a deep spectroscopic galaxy survey in fields around the quasar sightlines targeting nebular emission lines such as [\ion{O}{2}], [\ion{O}{3}], H$\beta$ and H$\alpha$ as well as stellar absorption features such as %\ion{Mg}{2}, \ion{Mg}{3}, \ion{Ca}{2}, and \ion{Fe}{2}
the 4000 \AA\ break, \ion{Ca}{2} H\&K, the G-band, \ion{Mg}{1}, and  \ion{Na}{1} D for precise redshifts using a `wedding cake' survey strategy. We observed each field with the Multi-Unit Spectroscopic Explorer (MUSE; \citealt{Bacon12MUSE}) on the Very Large Telescope (VLT), which can measure emission-line redshifts for galaxies as faint as $\approx 0.01L_*$ \citep[where $L_*$ refers to the characteristic rest-frame $r$-band absolute magnitude of $M_r=-21.5$ from ][]{Loveday12} at $z < 0.7$, out to $<30''$ around the quasar (corresponding to $d_{\rm proj} \approx 60-250$ kpc in within our redshift range). On intermediate scales, we used the Low Dispersion Survey Spectrograph 3 (LDSS3) on the Magellan Clay telescope using the VPH-ALL disperser, which can observe galaxies as faint as $\approx 0.1L_*$ at $z \approx 1$ to an angular radius of $3'$ (0.3-1.5 Mpc). On large scales, we performed galaxy spectroscopy with the Inamori-Magellan Areal Camera \citep[IMACS;][]{Dressler06IMACS} using the 200l and 150l dispersers to observe $L_*$ \new{galaxies} up to $z \approx 0.8$ to an angular radius of $10'$ (1-5 Mpc). The shallower but wider components of the CUBS galaxy survey enable us to identify and classify low-luminosity galaxies as field dwarfs or satellites of more luminous galaxies.

\begin{deluxetable*}{rrrrrrrrrr}
\tablecaption{Summary of galaxy properties in CUBS-Dwarfs \label{table:galaxyprops}}
\tablehead{
\colhead{Name} & \colhead{R.A.}& \colhead{Dec.} & \colhead{$d_{\rm proj}$} & \colhead{$R_{\rm vir}$} &
\colhead{$d_{\rm proj}$} & \colhead{$M_g$} & \colhead{$M_r$} & $\log M_{\star}$ & $\log M_{\rm h}$ \\
\colhead{} & \colhead{[J2000]}& \colhead{[J2000]} & \colhead{[kpc]} & \colhead{[kpc]} &
\colhead{$/R_{\rm vir}$} & \colhead{} & \colhead{} &   \colhead{$ / \rm{M_\odot}$} & \colhead{$ / \rm{M_\odot}$}
}

\startdata
J0154\_z0.12\_15  & $28.7294$  & $-7.2075$  & 15  & 90  & 0.17 & $-16.91$ & $-17.13$ & 8.3 & 10.8 \\
J0248\_z0.36\_16  & $42.0253$  & $-40.8099$ & 16  & 71  & 0.23 & $-17.53$ & $-17.89$ & 8.2 & 10.7 \\
J2339\_z0.08\_24  & $354.8113$ & $-55.4000$ & 24  & 47  & 0.52 & $-13.44$ & $-13.61$ & 6.8 & 9.9  \\
J2245\_z0.28\_42  & $341.2466$ & $-49.5306$ & 42  & 66  & 0.64 & $-14.92$ & $-15.26$ & 7.8 & 10.5 \\
J0333\_z0.16\_41  & $53.2741$  & $-41.0342$ & 41  & 61  & 0.68 & $-14.81$ & $-15.04$ & 7.5 & 10.3 \\
J2135\_z0.51\_64  & $323.9706$ & $-53.2794$ & 64  & 86  & 0.75 & $-17.14$ & $-17.49$ & 8.7 & 11.0 \\
J0333\_z0.20\_38  & $53.2793$ & $-41.0368$ & 38  & 47  & 0.81 & $-11.60$ & $-12.64$ & 7.2 & 10.1 \\
J0357\_z0.45\_67  & $59.3393$  & $-48.2072$ & 67  & 75  & 0.90 & $-17.38$ & $-17.41$ & 8.2 & 10.7 \\
\enddata
\tablecomments{The full table is available in the online version of the paper. \neww{We also include data from \cite{Johnson17}.}}
\end{deluxetable*}

We constrained the SED of each galaxy with $g$, $r$, $i$, $z$, $Y$ \new{photometry} from the Dark Energy Survey \citep{Abbott21DES2},  deeper $g$, $r$, $H$-band \new{photometry} from Magellan \citep{Li24CUBSVIII} \new{using IMACS/LDSS3 and FourStar \citep{Persson09FourStar} for optical and infrared respectively}, or synthetic broad-band photometry from the MUSE data \citep{CUBSI}. Following \cite{Johnson15}, we then inferred absolute rest-frame $g$- and $r$-band magnitudes using $k$-corrections estimated with the best fitting SED template from \cite{Coleman80}, and estimated the mass-to-light ratio using Equation \ref{eq:scaling} which is a fit to low-redshift galaxies in the NASA-Sloan Atlas \citep{Maller09SloanATLAS}. We then combined the mass-to-light ratio from Equation \ref{eq:scaling} with Equation \ref{eq:m2l} to infer stellar masses.

\begin{equation}
\small
\log \left(M_\star / L_r\right)= \begin{cases}-0.6 & M_g-M_r<0.15 \\ -1.01+2.95\left(M_g-M_r\right)  & M_g-M_r \geqslant 0.15 \\ -1.67\left(M_g-M_r\right)^2 & \end{cases}
\label{eq:scaling}
\end{equation}

\begin{equation}
\small
\log \left(M_\star / M_{\odot}\right)_{g r}=1.872-0.4 M_r+\log \left(M_\star / L_r\right)
\label{eq:m2l}
\end{equation}

These fitting functions recreate the stellar masses from the NASA-Sloan Atlas \citep{Maller09SloanATLAS} which were calculated using a \cite{Chabrier03imf} initial mass function, the kcorrect tool \citep{Blanton07kcorrect}, and GALEX+SDSS photometry. The resulting uncertainties in stellar masses are characterized by a  1$\sigma$ scatter of 0.15 dex. To infer corresponding halo masses, we used the stellar mass to halo mass relation from \cite{Kravtsov18}, and the virial radius as defined in \cite{Bryan98}. \neww{In one case (dwarf galaxy J0028\_z0.21\_117), the continuum is too faint to robustly constrain the stellar population, but emission lines are detected at high significance ($>5\sigma$). To infer a stellar mass for this dwarf, we measured an H$\alpha$-based star formation rate (SFR) using the relation  from \cite{Kennicutt12} and translated this to an expected stellar mass assuming the dwarf follows the mean relationship between SFR and stellar mass for low-mass, intermediate redshift galaxies from \cite{Boogaard18} (often referred to as the ``star-forming main sequence'').}

To create the CUB\neww{S}-Dwarfs sample from our deep and highly complete galaxy survey, we first identified \new{emission-line} galaxies of \new{$\log M_\star / {\rm M_\odot}< 9$}.
To select dwarf galaxies that are sufficiently isolated \new{from $\log M_\star / {\rm M_\odot}< 9$}, we then removed dwarfs that are within 500 km/s and 500 kpc projected distance of a neighboring galaxy with \new{$\log M_\star / {\rm M_\odot} > 9$}. \new{To remove systems with potential absorption from \new{luminous galaxies} close to the quasar sightline,} we also cut any dwarfs that have another galaxy of \new{$\log M_\star / {\rm M_\odot} > 9$} within 500 km/s of the dwarf galaxy redshift and 500 kpc of the quasar line-of-sight. To prevent confusion with any potential quasar outflows, we also removed galaxies within 10000 km/s of the quasar redshift. We further limited the redshift range of our sample to be $0.077 < z < 0.73$, with the upper and lower bound\new{s} reflecting redshifts for which the \ion{O}{6} $\lambda\lambda 1031, 1037{\rm \AA}$ doublet is in the observed COS spectral range.
    
\begin{deluxetable*}{lccllllll}
\tablecaption{Summary of total column densities of associated absorbers for each CUBS-Dwarfs galaxy \label{table:absorbers}}
\tablehead
{
\colhead{}&
  \multicolumn{2}{c}{}&
  \multicolumn{6}{c}{Column Density ($\log N / \rm cm^{-2}$)}
   \\
 \cline{4-9}
\colhead{Name}&\colhead{$z$}&\colhead{$d_{\rm proj} / R_{\rm vir}$}&  
\colhead{\ion{H}{1}}& 
\colhead{\ion{C}{2}}& 
\colhead{\ion{C}{3}}& 
\colhead{\ion{Si}{2}}& 
\colhead{\ion{Si}{3}}& 
\colhead{\ion{O}{6}}
}

\startdata
% J0154\_z0.12\_15 & 0.1163 & $0.17$ & $17.95_{-2.57}^{+0.73}$ & $14.04_{-0.02}^{+0.03}$ & $\hspace{-2.85mm}>\!14.73$ & $13.02_{-0.02}^{+0.01}$ & $\hspace{-2.85mm}<\!13.73_{-0.02}^{+0.02}$ & $14.18_{-0.01}^{+0.01}$ \\ 
% J0248\_z0.36\_16 & 0.3639 & $0.23$ & $17.57_{-0.09}^{+0.07}\hspace{3mm}$ & $14.58_{-0.09}^{+0.12}\hspace{3mm}$ & $14.12_{-0.06}^{+0.05}\hspace{3mm}$ & $12.53_{-0.19}^{+0.11}\hspace{3mm}$ & $13.96_{-0.10}^{+0.15}\hspace{3mm}$ & $14.76_{-0.06}^{+0.07}$ \\
J0154\_z0.12\_15 & 0.1163 & $0.17$ & $15.5 - 18.4$ & $13.98_{-0.01}^{+0.05}$ & $\hspace{-2.85mm}>\!14.73$ & $13.02_{-0.02}^{+0.01}$ & $\hspace{-2.85mm}<\!13.73_{-0.02}^{+0.02}$ & $14.18_{-0.01}^{+0.01}$ \\ 
J0248\_z0.36\_16 & 0.3639 & $0.23$ & $17.57_{-0.09}^{+0.07}\hspace{3mm}$ & $14.58_{-0.09}^{+0.12}\hspace{3mm}$ & $14.12_{-0.06}^{+0.05}\hspace{3mm}$ & $12.53_{-0.19}^{+0.11}\hspace{3mm}$ & $13.96_{-0.10}^{+0.15}\hspace{3mm}$ & $14.76_{-0.06}^{+0.07}$ \\
J2339\_z0.08\_24 & 0.0826 & $0.52$ & $15.0 - 18.1$ & $\hspace{-2.85mm}<\!13.15$ &  & $\hspace{-2.85mm}<\!12.06$ & $\hspace{-2.85mm}<\!12.29$ & $\hspace{-2.85mm}<\!13.66$ \\
J2245\_z0.28\_42 & 0.2757 & $0.64$ & $13.79_{-0.03}^{+0.03}$ & $\hspace{-2.85mm}<\!13.33$ & $\hspace{-2.85mm}<\!12.65$ & $\hspace{-2.85mm}<\!12.44$ & $\hspace{-2.85mm}<\!12.37$ & $\hspace{-2.85mm}<\!13.49$ \\
J0333\_z0.16\_41 & 0.1635 & $0.68$ & $14.49_{-0.22}^{+0.16}$ & $\hspace{-2.85mm}<\!12.93$ & $\hspace{-2.85mm}<\!13.46$ & $\hspace{-2.85mm}<\!12.20$ & $\hspace{-2.85mm}<\!11.91$ & $14.04_{-0.02}^{+0.02}$ \\
J2135\_z0.51\_64 & 0.5083 & $0.75$ & $14.29_{-0.03}^{+0.03}$ & $\hspace{-2.85mm}<\!13.52$ & $\hspace{-2.85mm}<\!13.83$ & $\hspace{-2.85mm}<\!12.99$ &  & $14.12_{-0.05}^{+0.06}$ \\
J0333\_z0.20\_38 & 0.2040 & $0.81$ & $\hspace{-2.85mm}<\!12.39$ & $\hspace{-2.85mm}<\!12.93$ & $\hspace{-2.85mm}<\!13.44$ & $\hspace{-2.85mm}<\!12.14$ & $\hspace{-2.85mm}<\!11.75$ & $\hspace{-2.85mm}<\!12.83$ \\
J0357\_z0.45\_67 & 0.4490 & $0.90$ & $14.60_{-0.01}^{+0.01}$ & $\hspace{-2.85mm}<\!13.21$ & $13.21_{-0.03}^{+0.02}$ & $\hspace{-2.85mm}<\!12.65$ & $\hspace{-2.85mm}<\!12.34$ & $14.39_{-0.01}^{+0.02}$ \\
\enddata
\tablecomments{The full table is available in the online version of the paper. \neww{We also include data from \cite{Johnson17}, though the upper limits for that data use the standard 2$\sigma$ method.}}
\end{deluxetable*}

For our analysis 
%\new{\st{going forward}}
we supplement our observations with the dwarf CGM sample from \cite{Johnson17}\new{,} which covered similar sets of UV absorption features. We also use the absorption fitting and galaxy stellar mass estimates for the LLS associated with a group of dwarf galaxies from Papers I/II\new{I} \citep{CUBSI, Zahedy21CUBSII}. CUBS-Dwarfs  \new{triples} the number of dwarf galaxies with \ion{H}{1} and \ion{O}{6} coverage within $d_{\rm proj}/R_{\rm vir} < 1$ when compared to \cite{Johnson17} from \neww{4} to \neww{12} \rm{in total (8 new)}. Moreover, within $1<d_{\rm proj}/R_{\rm vir}<3$, CUBS-Dwarfs increases the sample size by $\sim$11 times, from \neww{7} to \neww{80} in total \neww{(73 new)}. \neww{CUBS-Dwarfs also includes 10 galaxies at $d_{\rm proj} < 300$ kpc but $d_{\rm proj}/R_{\rm vir}>3$, bringing the total sample size of CUBS-Dwarfs to 91 dwarfs and 102 when systems from \cite{Johnson17} are included.} In Figure \ref{fig:sample}, we plot the redshift, stellar mass, and virial radius normalized projected distance distributions for the sample. In Table \ref{table:galaxyprops}, we report the stellar mass\new{es}, luminosit\new{ies}, and virial radius normalized projected distances ($d_{\rm proj}/R_{\rm vir}$) of the CUBS-Dwarfs sample. The sample has a stellar mass range of $\log M_\new{\star} / \rm{M_\odot} = 6.7 - 9.0$, with a median of $\log M_\new{\star} / \rm{M_\odot} = 8.3$, and median redshift of $z=0.3\new{1}$.

%still need to change these numbers

For comparison with previous work, the dwarf sample from \cite{Johnson17} covers a stellar mass range of $\log M_\new{*} / \rm{M_\odot} = 7.8 - 9.2$. At $z \approx 0$, the COS-Dwarfs survey \citep{Bordoloi14, Bordoloi18}, primarily observing \ion{C}{4} $\rm 1548\AA$ and $\rm 1550\AA$, sampled a higher range of $\log M_\new{*} / \rm{M_\odot} = 8.2 - 10.1$. Adding to the low-$z$ sample, \cite{Zheng23} observed a mass range of $\log M_\new{*} / \rm{M_\odot} = 6.5 - 9.5$.% At a higher redshift ($z > 0.1$), the CGM$^2$ survey \citep{Tchernyshyov22} has probed circumgalactic \ion{O}{6} at a wide range of masses, $\log M / M_\odot = 7.8 - 11$.

\subsection{Absorption-line measurements}
\label{sec:absorbers}

\begin{figure*}[t]
\centering
\includegraphics[width=0.90\textwidth]{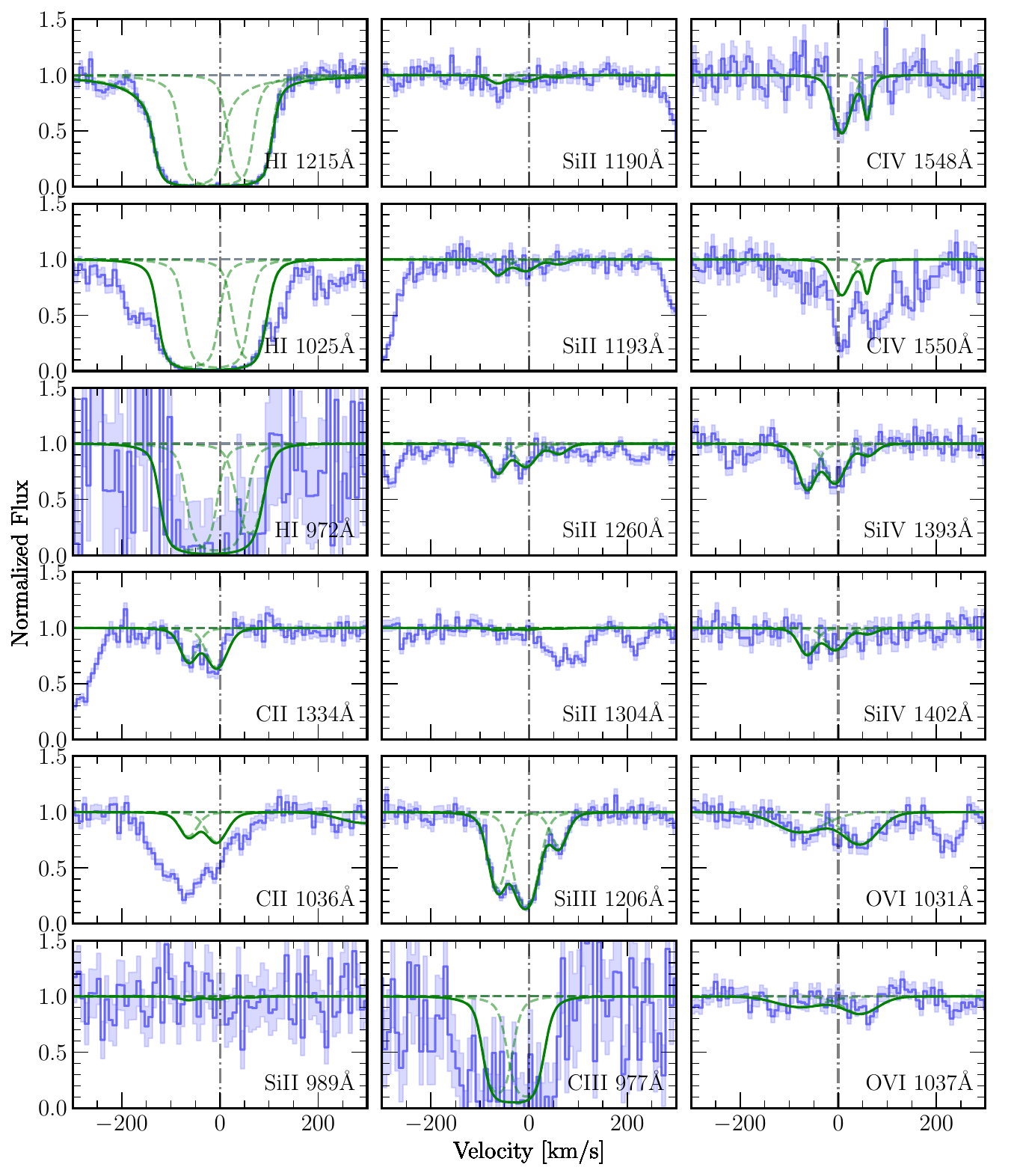}
\caption{Example Voigt profile fits for the CGM around a dwarf galaxy \new{ with a stellar mass of $\log M_\star / \rm{M_\odot} = 8.9$ at $z=0.1163$ and $d_{\rm proj} = 16$ kpc} \new{(ID: J0154\_z0.12\_15)}. The continuum-normalized quasar spectrum is plotted in blue histogram line-style with shaded bands indicating $1\sigma$ uncertainty after rebinning by a factor of three relative to the native pixel scale for display purposes. The best-fitting Voigt profile model is overplotted in a green solid line, with each individual component as dotted green lines. This example plots the system with the smallest virial radius normalized projected distance ($d_{\rm proj}/R_{\rm vir} \new{\approx 0.16}$) in the sample (aside from the LLS from \citealt{CUBSI}). We jointly fit the \ion{H}{1} and the lower ionization state metal ions and separately fit \ion{C}{4} and \ion{O}{6} due to their different kinematic structures. In cases of contamination (i.e. \ion{C}{2} $\rm 1036\AA$ or \ion{C}{4} $\rm 1550\AA$), we only fit the uncontaminated line and plot the predicted absorption over the contamination.} %Similar figures for each CUBS-Dwarfs galaxy are available on the journal webpage as a downloadable figure set.}
\label{fig:voigt}
\end{figure*}

We searched for absorbers within 300 km/s around the redshift of each dwarf galaxy. This search window exceeds the typical escape velocity at $d_{\rm proj}$ and the typical virial velocity of galaxies in CUBS-Dwarfs by a factor of $\gtrsim 2$,
assuming a Navarro-Frenk-White (NFW) profile \citep{Navarro96} with a concentration of 14 \citep{Diemer19}. To identify absorbers, we enforced a 3$\sigma$ detection threshold and followed up all potential detections by visual inspection and Voigt profile fitting to rule out contamination and artifacts.

For each galaxy with detected absorption, we measured the column density ($N$), Doppler width ($b$), and relative velocity to the galaxy ($\Delta v$) of \ion{H}{1}, \ion{C}{2}, \ion{C}{3}, \ion{C}{4}, \ion{Si}{2}, \ion{Si}{3}, \ion{Si}{4}, and \ion{O}{6} absorption with Voigt profile fitting of each individual absorber. \new{We performed the fitting using custom software built using AstroPy's \texttt{modeling} and \texttt{fitting} modules  \citep{Astropy13, Astropy18, Astropy22}.} This enables column density measurements while simultaneously mapping the kinematic structure of absorption. When possible, we jointly fit metal line absorbers with corresponding \ion{H}{1} absorbers using the thermal broadening ($b_{\rm therm}$) with the same temperature, the same non-thermal ($b_{\rm turb}$) broadening, and relative velocity ($\Delta v$). This is most often possible with low-ionization state metals that likely originate from gas at similar temperatures to \ion{H}{1}. However, the kinematic structure of higher ions such as \ion{C}{4} and \ion{O}{6} often differ from \ion{H}{1}, so we fit them separately.

We show an example Voigt profile fit in Figure \ref{fig:voigt}. To estimate uncertainties in column densities and other quantities inferred from the Voigt profile fitting, we ran Monte-Carlo Markov Chains \new{(MCMC; \citealt{ForemanMackey13})} to measure the posterior distributions of the fitted parameters. For non-detections, we report the 2$\sigma$ upper limit from \new{rest-frame} equivalent width ($W_r$) measured $\pm 45$ km/s around the redshift of the galaxy and converted to column density assuming the linear portion of the curve-of-growth. However, to account 
%\new{\st{for the}} 
for gas that may exist below the detection limit, we modified the traditional 2$\sigma$ upper limit estimate as:

\begin{equation}
    \mathrm{UL}(W_r) = \frac{W_r + |W_r|}{2} + 2\sigma_{W_r}
    \label{equation:limits}
\end{equation}

This accounts for any potential flux from a missed absorber but is identical to the classical $2\sigma$ upper limit ($2\sigma_{W_r}$) when the measured equivalent width is negative. \neww{To facilitate conversion between our limits and the 2-3$\sigma$ upper limits often adopted in the literature, we include $W_r$ and $\sigma_{W_r}$ values in the machine-readable tables accompanying this paper.} For ions with multiple observed transitions, we measured the upper limit for every available transition and reported the most stringent (lowest) value. The transitions with the highest oscillator strength typically provide the strongest limits. However, in cases of contamination, the upper limits can be sourced from lines with lower oscillator strengths. 

\section{Results}
\label{sec:results}

\begin{figure*}[t]
\includegraphics[width=\textwidth]{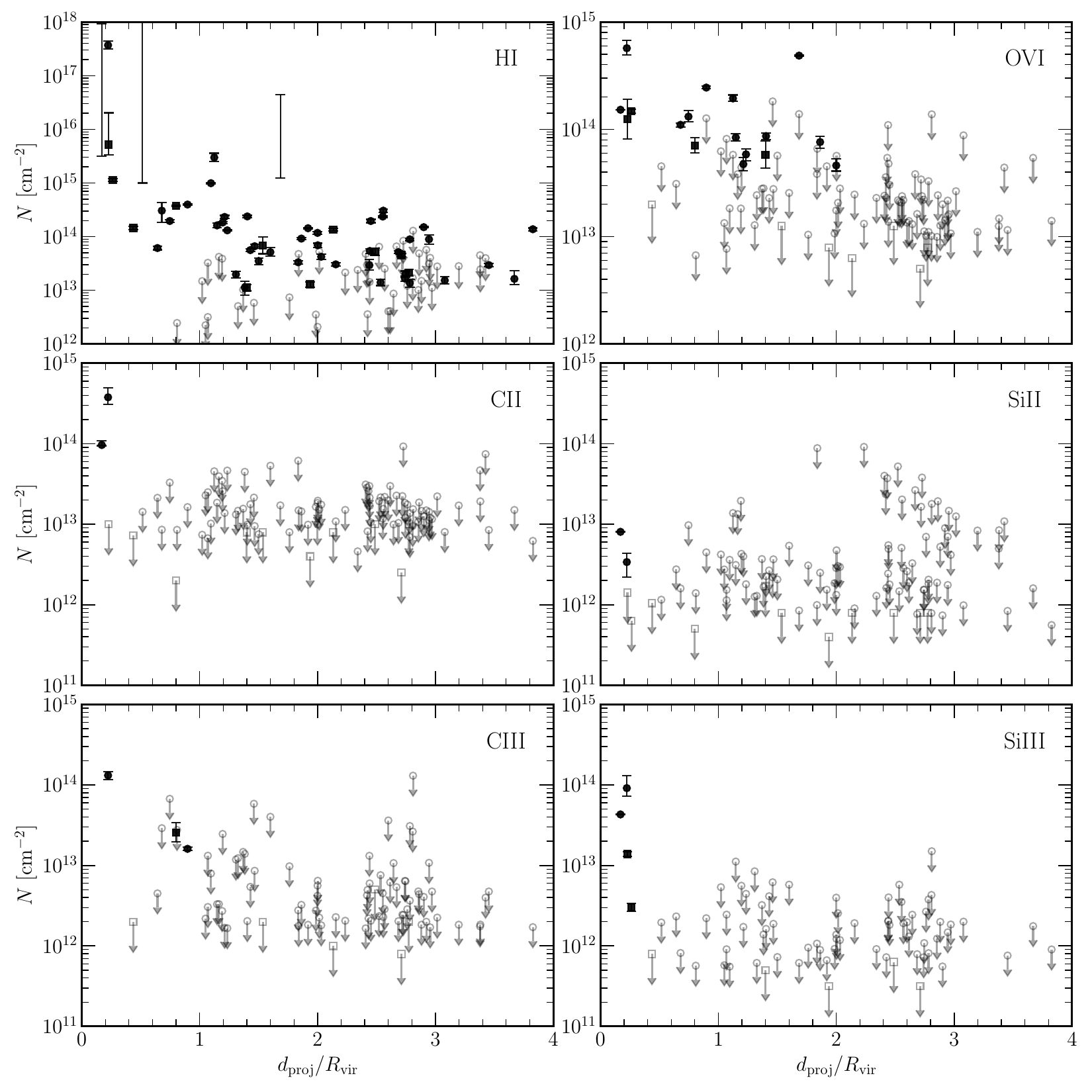}
\caption{Column densities ($N$) as a function of virial radius normalized projected distance ($d_{\rm proj}/R_{\rm vir}$) for \ion{H}{1}, \ion{O}{6}, \ion{C}{2}, \ion{C}{3}, \ion{Si}{2}, and \ion{Si}{3} as labeled in the top right of each panel. The sample from CUBS-Dwarfs is shown in circles and the column densities from \cite{Johnson17} are shown as squares. \new{Intervals with no points represent saturated systems which we provide upper and lower limits for. We color all \neww{the} points black to emphasize the uniformity of the sample selection across this work and \cite{Johnson17}.} The upper limits are estimated following Equation \ref{equation:limits} and are shown as open symbols with downward arrows. \new{\ion{H}{1} extends far out into the IGM with decreasing detection rate and column densities. Lower ionization state metals ions show a distinct lack of extended absorption, whereas highly ionized \ion{O}{6}, shows a high detection rate \new{at projected distances small than} the virial radius \new{and a non-negligible detection rate at $d_{\rm proj}/R_{\rm vir}=1-2$}.}} 
\label{fig:absorbers}
\end{figure*}
%
%%%%%%%%%%%%%%%%%%%%

To measure the strength of \ion{H}{1} and metal-line absorption as a function of distance from the dwarf galaxies, we report and visualize column densities (including upper limits) versus virial radius normalized impact parameter $d_{\rm proj}/R_{\rm vir}$, for \ion{H}{1}, \ion{O}{6}, \ion{Si}{2}, \ion{Si}{3}, \ion{C}{2}, and \ion{C}{3} in Table \ref{table:absorbers} and Figure \ref{fig:absorbers}. Normalizing the impact parameter by the virial radius allows us to compare absorption across the mass range of our sample, as recently demonstrated by \new{\cite{Qu23CUBSVI}}. 
%As demonstrated in \cite{Qu23CUBSVI}, utilzing $d_{\rm proj}/R_{\rm vir}$ more clearly demonstrates trends in the radial extent of gas for galaxy samples with a range of masses.
Within the 91 galaxy CUBS-Dwarfs sample, \neww{11 are coincident with another dwarf galaxy at a similar redshift}. For these, we associate the absorption to the dwarf with the smallest inferred 
%\new{\st{in}} 
$d_{\rm proj}/R_{\rm vir}$ \neww{and do not count their more distant dwarf members of the system towards the total number of galaxies in the CUBS-Dwarfs sample. As a check, we tested splitting the CUBS-Dwarfs sample into `fully isolated' (no detected neighbors) and `dwarf systems' (at least one neighboring dwarf galaxy) and did not find a statistically significant difference in covering fraction or average detected column density, though we caution that the sample of dwarf systems is small.} 

The upper left panel of Figure \ref{fig:absorbers} shows the relationship between the total column density of \ion{H}{1} systems (summing over systems with multiple absorption components within $\Delta v = \pm 300$ km/s) and $d_{\rm proj}/R_{\rm vir}$. 
The dwarf galaxies often exhibit strong \ion{H}{1} systems, such as \new{full} Lyman-limit systems (LLS, $\log N_{\rm HI} /\rm cm^{-2} > 17$) and partial Lyman-limit systems (pLLS, $\log N_{\rm HI} /\rm cm^{-2} > 16$), in sightlines that probe at closer projected distances. Restricting the sample to $d_{\rm proj}/R_{\rm vir} < 0.5$, we estimate the pLLS\new{+full LLS} covering fraction to be \new{$\kappa = 0.40_{-0.14}^{+0.22}$}. Moving outward, the pLLS\new{+full LLS} covering fraction within $0.5 < d_{\rm proj}/R_{\rm vir} < 1$ drops to \new{$\kappa = 0.14_{-0.05}^{+0.22}$}, though strong \ion{H}{1} absorption ($\log N_{\rm HI} /\rm cm^{-2} > 14$) remains common. Outside the virial radius, strong \ion{H}{1} absorption is rarer but detection rates remain non-zero even at $d_{\rm proj}/R_{\rm vir} \new{\approx} \new{4}$.% At higher redshift, the highest oscillator strength transition of \ion{H}{1}, Lyman-$\alpha$, moves out of the wavelength range of COS, which means we can only set upper limits using weaker lines. In some other cases, it is due to contamination in the stronger transitions which means the limit is again set by weaker lines.

The upper right panel of Figure \ref{fig:absorbers} displays detected column densities and upper limits for \ion{O}{6}. As with \ion{H}{1}, the covering fraction of strong \ion{O}{6} absorption is highly dependent on $d_{\rm proj} / R_{\rm vir}$. All systems with $d_{\rm proj} / R_{\rm vir} < 0.\new{4}$ show strong \ion{O}{6} absorption ($\log N_{\rm OVI} /\rm cm^{-2} > 14$), with the covering fraction decreasing to $\kappa = \new{0.38^{+0.13}_{-0.18}}$ for systems $ 0.4<d_{\rm proj} / R_{\rm vir}<1$. Outside the virial radius $1<d_{\rm proj}/R_{\rm vir}<2$, the covering fraction drops even further to $\kappa = 0.06_{-0.02}^{+0.07}$, and for $d_{\rm proj} / R_{\rm vir} > 2$ there are no detections within our sample, \new{setting a 2-$\sigma$ upper limit for the covering fraction of $\kappa < 0.03$}. %Similarly to the other ions, the detection fraction drops further away from the galaxy.
%CONTINUE HERE FOR CHANGES 06/11/24

However, in contrast to \ion{H}{1}, the \new{median} column density among detections remains relatively constant inside and outside the virial radius ($\log N_{\rm OVI} /\rm cm^{-2} \approx 14.1$ and $\log N_{\rm OVI} /\rm cm^{-2} \approx 13.\new{9}$, \new{with an intrinsic scatter of 0.3 dex for both}). This cannot be explained by spectral sensitivity, as we are sensitive to \ion{O}{6} column densities much lower than that, with the median \ion{O}{6} upper limit being $\log N_{\rm OVI} / \rm cm^2 \approx 13.\new{4}$. Therefore, while \new{the} mean \ion{O}{6} column density among detections is fairly flat, the mean averaged across all dwarfs, including \ion{O}{6} non-detections, decreases significantly with increasing $d_{\rm proj} / R_{\rm vir}$. The few instances of high upper limits originate from galaxies that have high levels of contamination at observed-frame wavelengths of both of the \ion{O}{6} 1031$\rm\, \AA$ and 1037$\rm \,\AA$ transitions.

The lower panels of Figure \ref{fig:absorbers} show the column densities for \ion{C}{2}, \ion{C}{3}, \ion{Si}{2}, and \ion{Si}{3}, which represent the other key metal lines commonly detected around massive galaxies and covered for the majority of the CUBS-Dwarfs sample. These ions have ionization potentials that are significantly lower than \ion{O}{6}, and typically probe cooler or denser gas phases. All of these metal lines have significantly lower detection rates than \ion{O}{6}, and no detections beyond $d_{\rm proj}/R_{\rm vir} \new{=} 1$. Considering silicon ions, detections of \ion{Si}{2} and \ion{Si}{3} are all limited to $d_{\rm proj}/R_{\rm vir} < 0.3$. The large differences in upper limits of \ion{Si}{2} are due to the wide range of oscillator strengths for 
%the $\lambda\lambda 990, 1021, 1190, 1193, 1260, 1304, 1526{\rm \AA}$ transitions.
the $\rm 98\new{9}\AA$, $\rm 102\new{0}\AA$, $\rm 1190\AA$, $\rm 1193\AA$, $\rm 1260\AA$, $\rm 1304\AA$, and $\rm 1526\AA$ transitions.
On the other hand, we find more carbon detections further out into the halo, with \new{two} \ion{C}{3} detections at $d_{\rm proj}/R_{\rm vir}=0.5-1$. \new{The lack of corresponding extended \ion{Si}{3} can be explained by the differing oscillator strengths of \ion{C}{3} and \ion{Si}{3} in conjunction with the abundances of carbon and silicon, as well the higher ionization potential of \ion{C}{3} (24.4 eV) compared to \ion{Si}{3} (16.3 eV).}
%and 1 \ion{C}{4} detection at $d_{\rm proj}/R_{\rm vir} > 1$. We also note that \ion{C}{4} $\rm 1548\AA$ and $\rm 1550\AA$ are only in the range of COS until $z \sim 0.3$, resulting in fewer measurements. Despite this, there are four \ion{C}{4} detections indicating a higher overall detection rate, with a covering fraction inside the virial radius of $\kappa = 0.30^{+0.16}_{-0.10}$. 
We note that \new{the} \ion{C}{4} $\lambda\lambda 1548, 1550{\rm \AA}$ and \ion{Si}{4} $\lambda\lambda 1393 {\rm \AA}, 1402 {\rm \AA}$ doublets are only in the range of COS \new{below} $z \new{\approx} 0.15$ and $z \new{\approx} 0.3$ respectively, resulting in fewer measurements. With this in mind, we leave the analysis of \ion{C}{4} and \ion{Si}{4} to future work with the addition of NUV spectra as part of the ongoing Circumgalactic Observations of NUV-shifted Transitions Across Cosmic Time (CONTACT; PID: 17517; PI: Chen) survey.

\begin{figure*}
\includegraphics[width=\textwidth]{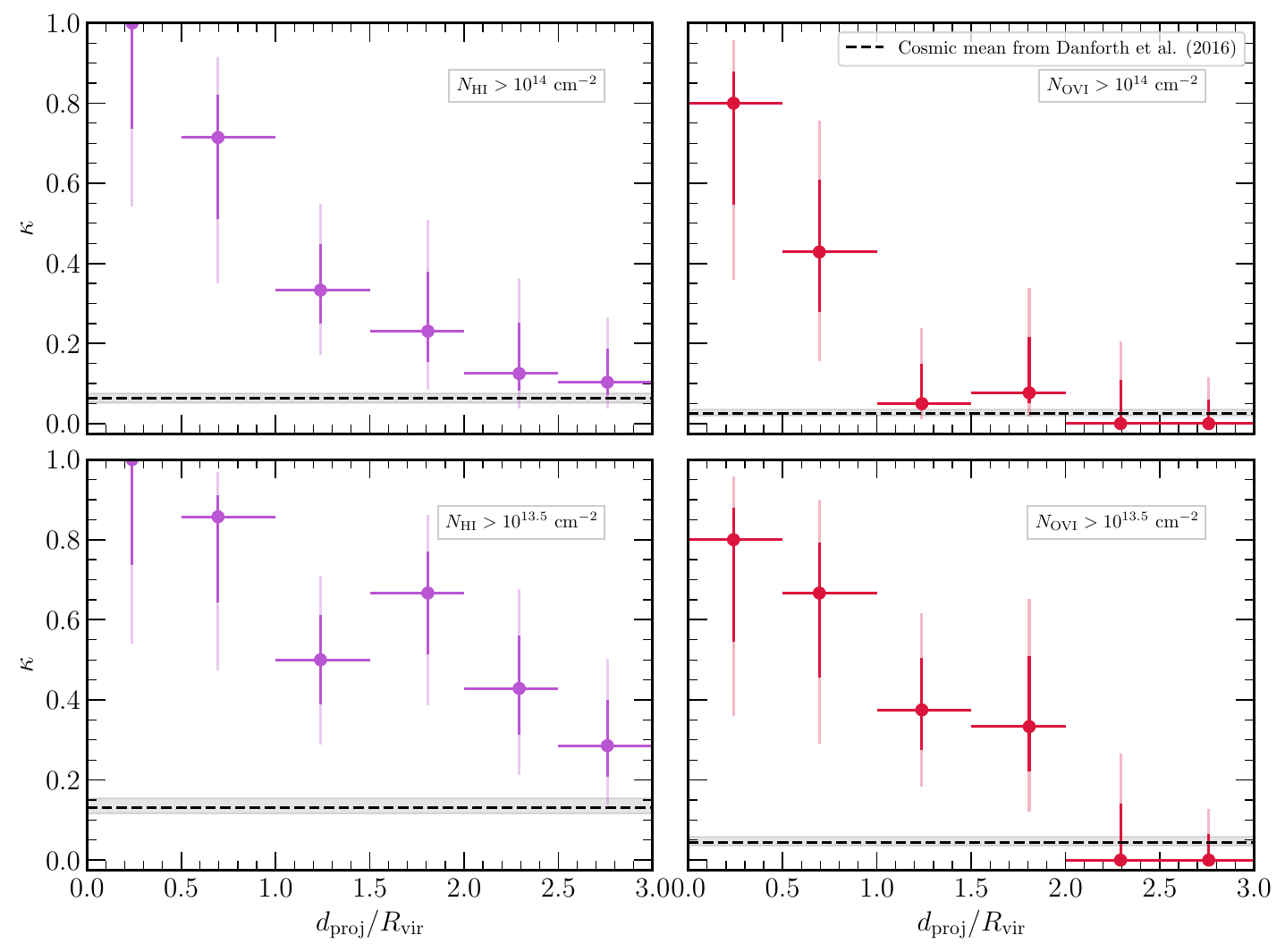}
\caption{Measured covering fraction\new{s} for \ion{H}{1} \new{({\it left panels})} and \ion{O}{6} \new{({\it right panels})} in 6 equal $d_{\rm proj}/R_{\rm vir}$ bins \new{for both \new{strong absorbers ($\log N/{\rm cm^{-2}}>14$; {\it top panels}) and weaker ones ($\log N/{\rm cm^{-2}}>13.5$; {\it bottom panels})}}. The dark and light vertical error bars mark the 68\% and 95\% confidence intervals respectively. The horizontal lines represent the $d_{\rm proj}/R_{\rm vir}$ range contained within each bin, with each point positioned at the median $d_{\rm proj}/R_{\rm vir}$ within the bin. We also include the covering fraction expected for random sightlines using the $z<0.7$ \ion{H}{1} and \ion{O}{6} column density distribution measured by \cite{Danforth16}. The grey bands around it show the 16th and 84th percentile redshifts from our survey. \new{The} \ion{H}{1} and \ion{O}{6} covering fraction\new{s} both decline as a function of \new{virial normalized impact parameter}.}% Notably, there is a non-negligible covering fraction outside the virial radius for \ion{O}{6}.} 
\label{fig:cf}
\end{figure*}

While low ionization metal lines are largely confined to the inner CGM of dwarf galaxies, strong \ion{H}{1} and highly ionized \ion{O}{6} are detected at larger distances. To quantify the relationship between strong ($\log N/{\rm cm^{-2}}>14$) \new{as well as weaker ($\log N/{\rm cm^{-2}}>13.5$)} \ion{H}{1} and \ion{O}{6} \new{absorption} as a function $d_{\rm proj}/R_{\rm vir}$, Figure \ref{fig:cf} displays their measured covering fraction in 6 bins. To compute the covering fraction, we first restrict the sample to those with sufficient S/N to ensure a confident \new{(3$\sigma$)} detection if the absorber strength is equal to the column density threshold. We then compute the fraction of detections above this threshold within this sample as well as the associated uncertainty following binomial statistics. %In Figure \ref{fig:cf}, we calculate a covering fraction by setting the detection threshold at $\log N_{\rm HI} /\rm cm^{-2} = 14$, following the convention from \cite{Johnson17}.
In the top left panel, we find the covering fraction for \ion{H}{1} systems with $\log N/{\rm cm^{-2}}>14$ at $d_{\rm proj}/R_{\rm vir} < 1$ is $\kappa = \new{0.83^{+0.06}_{-0.16}}$, dropping to $\kappa = \new{0.29^{+0.09}_{-0.07}}$ at $1<d_{\rm proj}/R_{\rm vir}<2$. The covering fraction continues to drop beyond $d_{\rm proj}/R_{\rm vir} > 2$, but never reaches zero. 
\new{The covering fractions for weaker \ion{H}{1} systems of $\log N/{\rm cm^{-2}}>13.5$ exhibit a similar declining trend, though with higher covering fractions, as expected with the lower column density threshold.}
The large sample size between 1 and 3 virial radii allows for relatively constraining estimates of $\kappa$ in these bins. To compare the measured covering fractions at large distance with expectations from the general IGM, the left panel of Figure \ref{fig:cf} displays the expected \ion{H}{1} detection fraction for systems with $\log N/{\rm cm^{-2}}>14$ \new{and $\log N/{\rm cm^{-2}}>13.5$} expected within $\pm 300\ \rm km\,s^{-1}$ in random sightlines assuming the low-redshift \ion{H}{1} column density distribution function from \cite{Danforth16}.
\new{While the \ion{H}{1} covering fractions decline with increasing virial radius normalized projected distance, they remain somewhat above the cosmic mean expected in random sightlines even at $d_{\rm proj}/R_{\rm vir} = 2.5-3$}.
%While the number of strong \ion{H}{1} absorbers is non-zero at large impact parameters, the covering fraction approaches the global mean expected of random sightlines.

The top right panel of Figure \ref{fig:cf}, displays the covering fraction for strong \ion{O}{6} ($\log N/{\rm cm^{-2}}>14$), which are below those of strong \ion{H}{1} at all projected distances in our sample. At $d_{\rm proj}/R_{\rm vir} < 1$ the dwarfs exhibit covering fractions of $\kappa = \new{0.58_{-0.14}^{+0.12}}$, which drops to $\kappa = \new{0.06_{-0.02}^{+0.07}}$ at $1<d_{\rm proj}/R_{\rm vir}<2$. Specifically, we identified two strong \ion{O}{6} detections at projected distances of $1<d_{\rm proj}/R_{\rm vir}<2$, implying a non-zero covering fraction. We find no detections of \ion{O}{6} at $d_{\rm proj}/R_{\rm vir}>2$, but the large number of galaxies in the bins allows for a stringent constraint on the covering fraction. Similar to the left panel\new{s}, we also plot the expected covering fraction for random sightlines assuming the \ion{O}{6} distribution function from \cite{Danforth16}. 
\new{The covering fractions for weaker \ion{O}{6} systems of $\log N/{\rm cm^{-2}}>13.5$ exhibit a similar declining trend with increasing virial radius normalized projected distance, though with higher covering fractions \neww{($\kappa=0.73^{+0.09}_{-0.16}$)} that remain well above the expected cosmic mean from random sightlines at $1<d_{\rm proj}/R_{\rm vir}<2$ \neww{($\kappa=0.36^{+0.10}_{-0.08}$)}.
With no  \ion{O}{6} detections at  $d_{\rm proj}> 2R_{\rm vir}$ however, the measured covering fractions for both strong and weaker systems drop to zero at $2<d_{\rm proj}/R_{\rm vir}<3$, consistent with expectations for random sightlines.} 

\neww{Comparing with the most recent study of the \ion{O}{6} absorption that includes low-mass galaxies, \cite{Tchernyshyov22} observed a higher covering fraction outside the virial radius than we found in CUBS-Dwarfs. Using their publicly available galaxy and absorption catalog, we compute an \ion{O}{6} covering fraction for galaxies with mass $\log M_\star/{\rm M_\odot} < 9$ of $\kappa = 0.27^{+0.16}_{-0.09}$ at $ 1<d_{\rm proj}/R_{\rm vir}<3$ for weak absorbers ($\log N/{\rm cm^{-2}}>13.5$) with CGM$^2$. This is  larger than our measured value of $\kappa = 0.15_{-0.03}^{+0.06}$, but is consistent within uncertainties. Some of this difference in covering fraction may be accounted for by differing distributions of $d_{\rm proj}/R_{\rm vir}$ between the two samples within the bin. However, we also note that we adopted significantly stricter isolation criteria to remove dwarfs that may be satellites of more massive galaxies. In particular, \cite{Tchernyshyov22}  removed dwarfs if there is a galaxy within a smaller $d_{\rm proj} / R_{\rm vir}$ within the 300 km/s search window, which can include satellites of massive galaxies. Indeed, our selection removes approximately half of the original CUBS galaxy sample with $\log M_\star / {\rm M_\odot} < 9$ due to proximity (within 500 kpc or 500 km/s) to a massive neighbor of $\log M_\star/M_\odot > 9$. Furthermore, both Figure 9 in \cite{Tchernyshyov22} as well as \cite{Johnson15}, demonstrate that \ion{O}{6} absorption is stronger and more extended in group environments, indicating that differences in the selection function described above may be important for explaining the tentative differences covering fraction. We also note that the CUBS survey is designed to reach higher spectroscopic completeness levels, with a typical completeness of 85\% for galaxies with observed i-band magnitudes brighter than 23 within 2 arcminutes of the quasar, versus $\sim 50\%$ for CGM$^2$ \citep{Wilde21CGM2}. However, we stress that the results of \cite{Tchernyshyov22} differ by only $\approx1\sigma$ from those reported here and that the difference is thus fully consistent with the expected statistical fluctuations.}

\neww{We also tested splitting the CUBS-Dwarfs sample into two mass bins along the median mass, and did not find a statistically significant difference in \ion{O}{6} covering fraction or average detected column density between the two samples, though the sample size in the two bins is limited. In future CUBS papers, we will compare the \ion{O}{6} covering fraction of dwarf galaxies to intermediate-mass ones.}

\section{Discussion}
\label{sec:discussion}

The column densities and covering fractions of highly ionized metal absorption around star-forming field dwarf galaxies in Figures \ref{fig:absorbers} and \ref{fig:cf} provide new constraints on the extent and state of the enriched CGM in low-mass systems.

\new{Before we interpret our results, we first qualify them with respect to potential projection effects along the line-of-sight. Using the EAGLE suite of simulations \citep{Schaye15EAGLE}, \cite{Ho21} showed that selecting absorbers by a velocity cut of $\pm 300$ km/s around simulated low-mass galaxies leads to a significantly higher covering fraction than computing a covering fraction using the true 3D distance selection of $d_{\rm proj}/R_{\rm vir}<1$ or even $d_{\rm proj}/R_{\rm vir}<2$. Our covering fraction for weak \ion{O}{6} matches their predicted value out to $d_{\rm proj}/ R_{\rm vir} = 2$ (See Figure 5 in \citealt{Ho21} vs. Figure \ref{fig:cf} in this work). However, our results diverge at $2< d_{\rm proj}/ R_{\rm vir}<3$, with \cite{Ho21} predicting $\kappa_{\rm OVI}(\log N/{\rm cm^{-2}}>13.5)  \approx 0.2$, whereas we found no such \ion{O}{6} absorbers at $d_{\rm proj}/ R_{\rm vir}=2-3$ and set a 2$\sigma$ upper limit at $\kappa_{\rm OVI}(\log N/{\rm cm^{-2}}>13.5)  \lesssim 0.1$. These results suggest that while a non-negligible fraction of the \ion{O}{6} absorption observed around the dwarf galaxies in this work may arise at $1 < d_{\rm proj}/ R_{\rm vir} < 2$, relatively little is coming from larger 3D distances.}

\new{With this caveat in mind}, we discuss our results in two subsections. First, we consider the implications of the observed \ion{O}{6} covering fractions for the metal and baryon mass accounting in the CGM and IGM around isolated, star-forming dwarf galaxies. Second, we discuss the kinematic distribution of the metal ions and quantify the fraction of observed metals that are not bound to the dwarfs.

\subsection{Column density and ion mass}

The CGM and IGM around the CUBS-Dwarfs \new{galaxies} exhibit low detection rates of low and intermediate metal ion absorption from silicon and carbon ions (\ion{Si}{2}, \ion{Si}{3}, \ion{C}{2}, \ion{C}{3}). This is consistent with previous studies showing that the cool photoionized CGM and IGM around dwarfs constitute only a modest fraction of the metal and baryon budgets of the systems \citep[e.g.,][]{Johnson17, Zheng23}. 
\new{Among higher ions}, we found comparatively high detection rates of \ion{O}{6} absorption, \new{consistent with \cite{Johnson17}}  \citep[also see consistent results from][in review]{MUSEqUBESOVIMassDutta24}.. \new{Although both analytic models and simulations have shown that the virialization/heating of the CGM is non-uniform \citep[e.g.,][]{Stern19, Stern20},} the galaxies in the CUBS-Dwarfs sample fall below the mass thresholds for stable hot halo formation \citep[\new{e.g.,}][]{Correa18}. Nonetheless, the origins of the \ion{O}{6} absorption around the dwarf galaxies are therefore likely qualitatively different than around more massive galaxies, and may involve a combination of collisional ionization and photoionization, or possibly non-equilibrium conditions. New NUV spectra enabling combined constraints on both \ion{O}{6} and \ion{C}{4} are needed for robust insights into the physical conditions of the \ion{O}{6} absorbing gas.

\begin{figure}[t]
\includegraphics[width=0.49\textwidth]{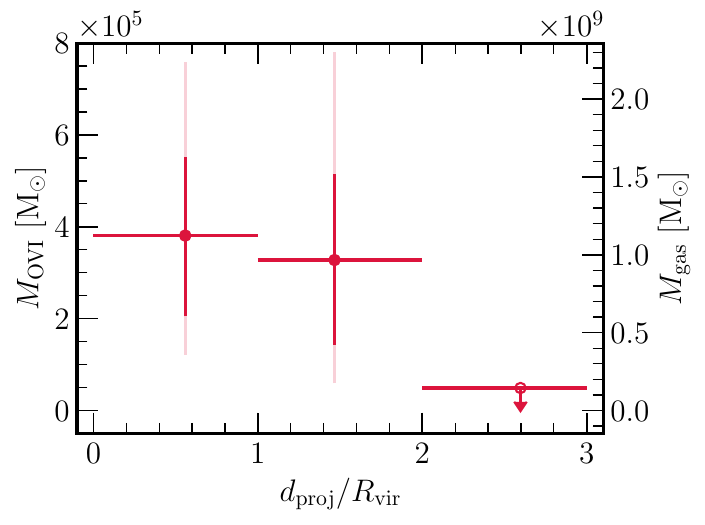}
\caption{\new{Left} axis: The \ion{O}{6} masses inferred from the covering fractions in Figure \ref{fig:cf} and $\langle N_{\rm OVI} \rangle$ above the cutoff, using Equation \ref{eq:mass}. \new{The vertical red and light red bars represent the 1$\sigma$ and 2$\sigma$ errors inferred from bootstrapping}. We find that a non-negligible fraction of the ion mass is outside the virial radius. \neww{The point in the third bin is the mass inferred from coadding the spectra from non-detections to set an overall mass upper limit in the outer annulus.} \new{Right} axis: The inferred total gas mass within each bin assuming an ionization fraction $f \approx 0.2$ and a metallicity of $0.3Z_\odot$.}

\label{fig:ionmass}
\end{figure}

The \new{spatial} distribution of \ion{O}{6} \new{around} CUBS-Dwarfs \new{galaxies} differs from \ion{H}{1} as the average column density among detections remains relatively constant as a function of virial normalized projected distance, as shown in Figure \ref{fig:absorbers}. However, this average is dominated by a few strong absorbers, and \ion{O}{6} absorbers are rare in the outer CGM, but can have high column densities when detected. As a result, they can therefore still represent a significant fraction of the total \ion{O}{6} mass in the CGM and IGM around dwarf galaxies. 
To estimate the oxygen mass detected in \ion{O}{6} within the CGM and IGM around dwarf galaxies, we adopt the median galaxy mass in CUBS-Dwarfs ($\log M_\star/\rm{M_\odot} \new{\approx} 8.3)$, and approximate the mass in annuli following,

\begin{equation}
    M_{\rm ion} \approx \pi  (R_{\rm outer}^2 - R_{\rm inner}^2) m_{\rm ion} \kappa_{\rm ion} \langle N_{\rm ion} \rangle,
    \label{eq:mass}
\end{equation}

where $m_{\rm ion}$ is the mass of the ion, $\kappa_{\rm ion}$ is the covering fraction of the ion measured in the annulus, $\langle N_{\rm ion} \rangle$ is the mean column density among detections within the annulus, and $R_{\rm in\neww{ner}}$/$R_{\rm out\neww{er}}$ are the inner/outer radii of the annulus. \new{We performed this calculation twice, both with strong ($\log N/{\rm cm^{-2}}>14$) and weak ($13.5<\log N/{\rm cm^{-2}}<14$) detections and summed the resulting masses.}
In Figure \ref{fig:ionmass}, we show $M_{\rm ion}$ in two annuli, from $R_{\rm inner}=0$ to $R_{\rm outer}=R_{\rm vir}$, and $R_{\rm inner}=R_{\rm vir}$ to $R_{\rm outer}=2R_{\rm vir}$, roughly representing the CGM and nearby IGM around the dwarf galaxies. We find $M_{\rm OVI} \new{=} \new{3.6^{+1.6}_{-1.6}} \times 10^5 \rm\, M_\odot$ and $\new{3.3^{+1.8}_{-1.9}} \times 10^5 \rm M_\odot$ in the two bins, respectively. \neww{To calculate the potential mass falling below the detection limit, we co-add the spectra with column density upper limits below $\log N/{\rm cm^{-2}}<13.5$ using inverse variance weighting. We find an upper limit on the average column density of $\log N/{\rm cm^{-2}} < 12.4$ in the $d_{\rm proj}/R_{\rm vir}=2-3$ bin. This corresponds to a mass upper limit of $M_{\rm O\,VI} < 3.4 \times 10^4 \rm\, M_\odot$ which we plot in Figure \ref{fig:ionmass} as an open circle with a downward arrow. The non-detection and upper limit implies that the \ion{O}{6} mass associated with star-forming field dwarf galaxies is dominated by the inner two annuli at projected distances of $d_{\rm proj}/R_{\rm vir}<2$.}

%Stars and the ISM account for 20\% of expected metals created from star-formation in sub-$L_*$ galaxies \citep{Peeples2014}, implying that a significant fraction of metals must have been lost to the CGM or IGM via outflows \citep[e.g.,][]{Tremonti2004}. \ion{O}{6} is one of the most common metal ions detected in the CGM and is, therefore, a potential tracer of this process. 
%To contextualize our estimated \ion{O}{6} ion mass with the expected baryon budget, we compute the total CGM mass using values for the ionization fraction ($f \approx 0.2$) and solar oxygen abundance ($n_{\rm O}/n_{\rm H} \approx 5 \times 10^{-4}$) consistent with \cite{Tumlinson11}, with gas metallicity ($Z \approx 0.3 Z_\odot$) consistent with assumptions in \cite{Zheng23}. 

To contextualize our estimated \ion{O}{6} mass, we infer the fraction of the oxygen budget in the \ion{O}{6}-bearing phase of the CGM assuming an ion fraction of $f = 0.2$. The \ion{O}{6} ion fraction is not expected to exceed $\approx 0.2$ under collisional or photoionization equilibrium conditions \citep{Oppenheimer13} and is consistent with the assumptions made in \cite{Tumlinson11} and \cite{Johnson17}. 
Adopting an IMF-normalized oxygen yield of $0.015 \rm{M_\odot}$ per solar mass of star formation \citep{Peeples2014} \new{as was assumed in previous CGM studies \citep[e.g.,][]{Tumlinson11, Johnson17, Rudie19KBSS},} we find that the inferred oxygen mass in the \ion{O}{6}-bearing phase represents $\approx \new{9}0\%$ of the total oxygen budget expected from supernova yields associated with past star formation in the dwarfs. \new{We note, however, that IMF-weighted oxygen yields of stellar populations are highly uncertain, due to uncertainties in the IMF shape and the nucleosynthesis contributions of the most massive stars \citep{Vincenzo16}. Reasonable assumptions about these processes lead to yields that range from 0.006 $\rm{M_\odot}$ \citep{Nomoto06} to $>0.03\, \rm{M_\odot}$ \citep{Vincenzo16} of oxygen per solar mass of star formation. These uncertainties, however, do not preclude comparisons to other observed or simulated CGM samples as long \neww{as} yield assumptions are consistent.}

We also use the \ion{O}{6} mass to estimate the total gas mass $M_{\rm gas}$ of the \ion{O}{6}-bearing CGM/IGM around the dwarfs using solar abundance patterns \citep{Asplund09} scaled to a metallicity of $Z \approx 0.3\, Z_\odot$ to be consistent with the analysis in \cite{Zheng23} \new{as well as estimates for the metallicity for \ion{O}{6}-bearing gas from the EAGLE simulations \citep[][]{Rahmati16}}.  With this and our previous ion fraction assumption, we calculate $M_{\rm gas} = \new{1.1^{+0.5}_{-0.5}} \times 10^9 \rm \, M_\odot$ at $d_{\rm proj}/R_{\rm vir}<1$ and $\new{9.6^{+5.4}_{-5.4}} \times 10^8 \rm\, \, M_\odot$ at $1<d_{\rm proj}/R_{\rm vir}<2$. \new{We include the total gas mass estimate in Equation \ref{eq:mass} to enable readers to scale to their choice of ionization fraction and/or metallicities}:

\begin{equation}
    \new{M_{\rm gas} = 2.1_{-0.7}^{+0.7} \times 10^9 \, {\rm M_\odot} \, \left( \frac{0.3 \, {\rm Z_\odot}}{Z} \right)  \left( \frac{0.2}{f} \right)}
    \label{eq:mass}
\end{equation}

Therefore, we estimate the gas mass in the CGM of the galaxies in CUBS-Dwarfs ($\log M_{\rm gas}/\rm{M_\odot} \approx 9.
\new{3}$) to be larger than the median stellar mass of the sample ($\log M_\star/\rm{M_\odot} \approx 8.3$) and typical ISM \ion{H}{1} mass detected in 21-cm emission for local star-forming dwarf galaxies of similar stellar mass \citep[\new{$\log{M_{\rm HI,\,21 \,cm} / \rm M_\odot}\approx 8.9$;} e.g.,][]{Parkash18}.
%\new{\st{Comparing to the total baryon budget, $M_{\rm b} = M_{\rm halo} \Omega_b$, we find that the CGM represents $\gtrsim13\%$ of the baryon budget of a typical CUBS-Dwarfs galaxy.}
We also note that the total oxygen mass and gas mass at $d<R_{\rm vir}$ found around the dwarfs \neww{is comparable to what \cite{Tumlinson11}} found around $L_*$ galaxies despite the factor of 2 dex difference in median stellar mass. This suggests that the \ion{O}{6}-\new{bearing gas represents a larger} fraction of the mass of the CGM \new{around} dwarf galaxies \new{than around} $L_*$ galaxies.
%This implies that a significant fraction of metals must have been lost to the CGM or IGM via outflows \citep[e.g.,][]{Tremonti2004}, with \ion{O}{6}-enriched gas acting as a potential tracer of this material \citep[e.g.,][]{Tumlinson11, Muzahid15, Rosenwasser18, Grimes09}. 
%We show that a significant fraction of \ion{O}{6} in dwarf galaxies has been ejected past the virial radius. 
%Using a typical value for ionization fraction, $f \approx 0.2$, solar oxygen abundance ($n_{\rm O}/n_{\rm H} \approx 5 \times 10^{-5}$), and $0.3 Z_\odot$ gas metallicity, we find $M_{\rm gas} \approx 1.8 \times 10^9 \rm M_\odot$ within the virial radius and $M_{\rm gas} = 4 \times 10^8 \rm M_\odot$ outside the virial radius. Therefore, we estimate the gas mass in the CGM of the galaxies in CUBS-Dwarfs ($\log M_{\rm gas}/M_\odot \approx 9.2$) to be significantly larger than the median stellar mass of the sample ($\log M_\star/M_\odot \approx 8.4$).

\subsection{Kinematics}

\begin{figure}[t]
\includegraphics[width=0.48\textwidth]{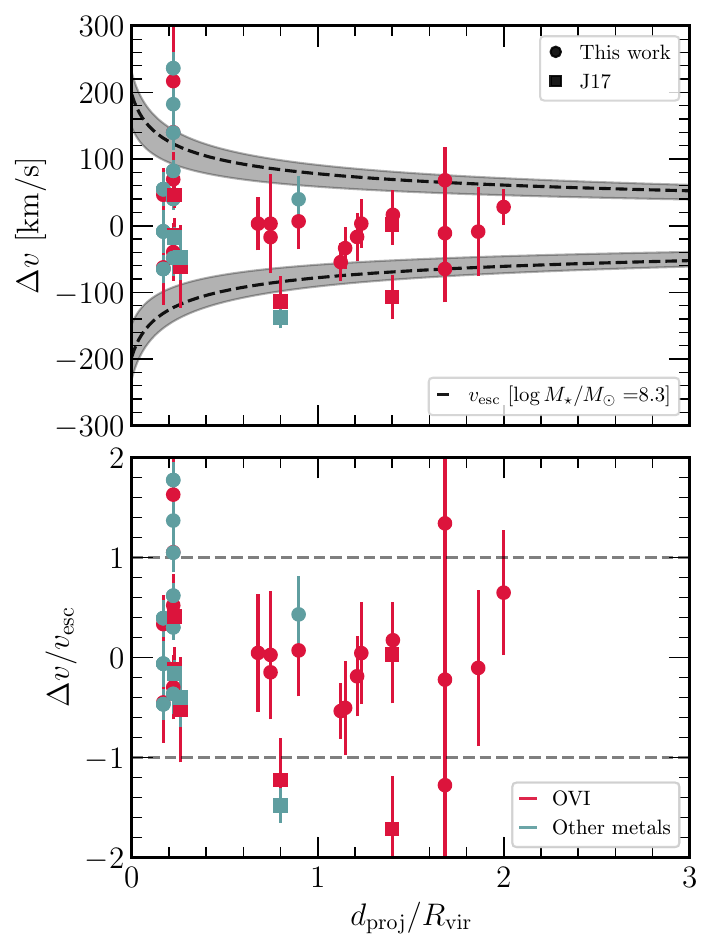}
\caption{\new{Top: }The radial velocity of metal absorption components relative to the dwarf galaxy systemic velocity versus virial radius normalized projected distance ($d_{\rm proj}/R_{\rm vir}$). The error bars represent $\pm b$, the Doppler width of each component. \ion{O}{6} components are plotted \new{as} red points while other metal lines are shown in teal. We also show curves for the escape velocity of an NFW profile for the median mass of our sample. The grey bands show the same curve for the 16th and 84th mass percentile in the CUBS-Dwarfs sample. \new{Bottom: The radial velocity of absorbers, normalized by the estimated escape velocity of associated galaxy halos. The dotted lines represent the border between the formally bound and unbound absorbers. }Approximately \neww{$15\%$} of the dwarf galaxies exhibit \new{formally} unbound metal absorption.}

\label{fig:kinematics}
\end{figure}

The kinematics of the circumgalactic medium of dwarf galaxies can provide insights into the outflows thought to regulate galaxy evolution. % To infer the fraction of ion mass being ejected by feedback mechanisms we need to calculate the fraction of absorbers that are moving beyond $v_{\rm esc}$ of the halo. 
However, because we can only observe the line-of-sight (LOS) velocity of the gas, we cannot infer the true radial velocity of the gas with respect to the galaxy due to projection effects, complicating the interpretation of absorbing gas as outflowing vs inflowing. Therefore, we can only unambiguously show that gas is unbound to the halo if the LOS velocity is greater than \new{the escape velocity} ($v_{\rm esc}$) as shown in Figure \ref{fig:kinematics}. We note that some of the other absorbers could also be unbound if their transverse velocities relative to the dwarf or true 3D distance from the dwarf are sufficiently high. Each point \new{in the top panel of} Figure \ref{fig:kinematics} shows an absorber at a given $d_{\rm proj}/R_{\rm vir}$ and the corresponding LOS velocity inferred from the Voigt profile fit. The `error bars' shown in the figure are the Doppler width ($b$) of the absorber. With dotted lines, we show the 1D escape velocity inferred from an NFW profile of the median galaxy in stellar mass ($\log M_\star/\rm{M_\odot} \approx 8.3$) using the stellar-mass halo mass relation from \cite{Kravtsov18}, the concentration halo mass relation from \cite{Diemer19}, and the Colossus package introduced in \cite{Diemer18} to generate the escape velocity profile. \new{Similarly, the bottom panel shows the relative velocity of each absorber normalized by the estimated escape velocity of the associated galaxy halos.}

Following the methodology of \cite{Rudie19KBSS}, we calculate the fraction of dwarf galaxies that have an associated metal absorber with a LOS velocity greater than $v_{\rm esc}$ at the corresponding $d_{\rm proj}$ for the galaxy's halo mass. We find that 4/26 ($\approx15\new{^{+10}_{-5}}\%$) dwarfs with detected metal line absorption have at least 1 unbound absorber. For comparison, \cite{Tumlinson11}  found that 4/27 ($\approx15\new{^{+10}_{-5}}\%$) star-forming galaxies \new{at $0.1<z<0.36$ with stellar masses $9.5 < \log M_\star/\rm{M_\odot} < 11.5$ have \ion{O}{6}-bearing absorbers that are formally unbound from their halos  \citep[also see consistent results from ][in review]{MUSEqUBESOVIKinematicsDutta24}.}  In contrast, \cite{Rudie19KBSS} studied galaxies at $2.1 < z < 2.7$ with a mass range of $9 < \log M_\star/ \rm{M_\odot} < 10.7$ and found the fraction of galaxies with  \new{formally} unbound metal line absorbers to be $\new{\approx} 7\new{0}\%$. This difference implies that the IGM/CGM traced by ionized metal absorbers has a significantly smaller unbound fraction at $z<1$ versus $z>2$.

While the fraction of dwarf galaxies with detected unbound \ion{O}{6} lines is low, Figure \ref{fig:absorbers} demonstrates that the \ion{O}{6} absorbers in our sample have a wide range of column densities. Therefore, to estimate the fraction of $M_{\rm OVI}$ that is unbound, we need to calculate the fraction of total column density that is unbound within a given absorber. \new{We compute this quantity using two methods. First, we calculated the column density weighted fraction of absorbers that are formally unbound based on the relative velocity of the centroid of each absorber. Second,} we measure the inferred column density from the optical depth (as long as the absorption is unsaturated, which all our \ion{O}{6} absorption is) of each pixel of an absorber from the Voigt profile fit, measuring its relative velocity to determine whether it is formally unbound. This means that absorbers with large Doppler widths $b$ and 
%\new{\st{have}} 
LOS velocities close to $v_{\rm esc}$ will be partially unbound. \new{Using both methods and given our assumptions, we find that at least $\approx\neww{15}\%$} of the \ion{O}{6}-bearing gas within the halos of CUBS-Dwarfs galaxies is \new{formally} unbound, with the fraction being roughly equal both inside and outside the virial radius. \new{However, we caution that there are potentially large modeling uncertainties in the estimated escape velocity profiles of dwarf galaxies due to uncertainty in the stellar-to-halo mass relation and dark matter profiles in low mass systems \citep[see discussion in][]{Thornton23}.}

\section{Summary and Conclusions}
\label{sec:summary}

In this study, we presented a comprehensive analysis of the multi-phase circumgalactic medium (CGM) and intergalactic medium (IGM) in the vicinity of star-forming field dwarf galaxies using quasar absorption line measurements from sightlines passing near \neww{91} dwarfs discovered in CUBS. Our examination of the CGM and IGM around these dwarf galaxies provides new insights into the distribution and kinematics of species including \ion{H}{1} and \ion{O}{6}, as well as \ion{C}{2}, \ion{C}{3}, \ion{Si}{2}, and \ion{Si}{3}, thereby improving our understanding of the baryonic processes at work around these low-mass galaxies. Our primary findings from this work are as follows:

\begin{itemize}
    \item The average detected column density and covering fraction of strong \ion{H}{1} absorbers ($\log N_{\rm HI} /{\rm cm^{-2}}>14$) declines with virial normalized projected distance, with  $\kappa_{\rm HI} = \new{1.00_{-0.17}^{+0.00}}$ at projected distances less than half of the virial radius, and steadily dropping to $\kappa_{\rm HI} = \new{0.11_{-0.03}^{+0.06}}$ at projected distances of \new{$2-3$} times the virial radius \new{(see Figure \ref{fig:absorbers})}.
    
    \item Low ionization metal absorption traced by \ion{C}{2} and \ion{Si}{2} is primarily confined to the inner CGM with no detections at projected distances greater than half the estimated halo virial radii \new{(see Figure \ref{fig:absorbers})}.
    
    \item Intermediate ionization absorption traced by \ion{C}{3} is more extended than \ion{C}{2}, but with detections confined to projected distances smaller than the virial radius \new{(see Figure \ref{fig:absorbers}).}
    
    \item \neww{In contrast with low and intermediate ions, highly ionized \ion{O}{6} is commonly detected at $d_{\rm proj}/R_{\rm vir} <1$ and} shows a distinct decline in covering fraction for strong absorbers ($\log N_{\rm OVI} /{\rm cm^{-2}}>14$) at increased projected distances\new{,} with $\kappa_{\rm OVI} = \new{0.50_{-0.13}^{+0.13}}$ at projected distances within the virial radius, $\kappa_{\rm OVI} = 0.06_{-0.02}^{+0.07}$ at projected distances of $1-2$ times the virial radius, and no detections at larger projected distances. \neww{For weaker absorbers ($\log N_{\rm OVI} /{\rm cm^{-2}}>13.5$), the covering fraction is higher inside the virial radius, with $\kappa_{\rm OVI}=0.77^{+0.08}_{-0.15}$, and remains well above the cosmic mean at $1<d_{\rm proj}/R_{\rm vir} <2$ with $\kappa_{\rm OVI}=0.14^{+0.06}_{-0.03}$}. Interestingly, while the covering fraction decreases, the mean column density among detections remains fairly flat \new{(see Figures \ref{fig:absorbers} and \ref{fig:cf})}.
    
    \item With minimal ionization corrections, the oxygen contained within the \ion{O}{6}-bearing phase of the CGM and IGM nearby dwarf galaxies represents a substantial fraction of the metal budget ($\new{\approx} 90\%$) associated with supernova yields expected from past star formation. This suggests that a large fraction of metals produced by supernovae explosions in dwarf galaxies and expelled out of their ISM remain at relatively modest distances within the nearby CGM and IGM at $\lesssim 2 R_{\rm vir}$.
    
    \item Adopting this minimal ionization correction\new{,} and assuming a metallicity of $Z=0.3\,Z_\odot$, the \ion{O}{6}-bearing phase of the CGM/IGM at $\lesssim 2 R_{\rm vir}$ from the dwarfs carries a total gas mass of at least $M_{\rm gas} \approx 2 \times 10^9 \, \rm M_\odot$ (see Figure \ref{fig:ionmass}).
    %a significant fraction of the ion mass may be situated outside of the virial radius of dwarf galaxies, indicating that metal enrichment of the IGM from dwarf galaxy feedback processes may be an important contributor to the metal content seen in the IGM.
    \item The kinematic data suggest that a modest $\approx \neww{15}\%$  of the detected \ion{O}{6} gas, is likely unbound from the dwarf galaxies \new{(see Figure \ref{fig:kinematics}), comparable to what is found around Milky Way-like galaxies.}
    %This result is consistent with that of more massive galaxies \cite{Tumlinson11}, but contrasts with higher redshift ($z \sim 2$) galaxies in \cite{Rudie19KBSS} which found a significantly larger unbound fraction.
\end{itemize}

\section*{Acknowledgements}

We thank the referee for taking the time to review our work and providing detailed and insightful comments. We thank S. Danieli and E. Bell for discussion of different definitions of dwarf galaxies. This research is based on observations made with the NASA/ESA Hubble Space Telescope obtained from the Space Telescope Science Institute, which is operated by the Association of Universities for Research in Astronomy, Inc., under NASA contract NAS 5–26555. These observations are associated with program: GO-15163. S.D.J. acknowledges partial support from STScI grants HST-GO-16728.001-A, HST-GO-15935.021-A, and HST-GO-15655.018-A. H.W.C., Z.Q., and M.C. acknowledge partial support from NSF AST-1715692, HST-GO-15163.01A, and NASA ADAP 80NSSC23K0479 grants. E.B. acknowledges support by NASA under award No. 80GSFC21M0002. C.A.F.G. was supported by NSF through grants AST-2108230, AST-2307327, and CAREER award AST-1652522; by NASA through grants 17-ATP17-0067 and 21-ATP21-0036; by STScI through grants HST-GO-16730.016-A and JWST-AR-03252.001-A; and by CXO through grant TM2-23005X. J.I.L. is supported by the Eric and Wendy Schmidt AI in Science Postdoctoral Fellowship, a Schmidt Futures program. \new{S.L. acknowledges support by FONDECYT grant 1231187.} Based
on observations collected at the European Organisation for Astronomical Research in the Southern Hemisphere under ESO program 104.A-0147. We would analysis in this work was aided by the following free software packages, \texttt{Matplotlib} \citep{matplotlib}, \texttt{NumPy} \citep{numpy},  \texttt{SciPy} \citep{scipy}, and \texttt{Astropy} \citep{Astropy13, Astropy18, Astropy22}.  This paper includes data gathered with the 6.5-meter Magellan Telescopes located at Las Campanas Observatory, Chile. This research has made use of NASA's Astrophysics Data System Bibliographic Services. Some/all of the data presented in this article were obtained from the Mikulski Archive for Space Telescopes (MAST) at the Space Telescope Science Institute. The specific observations analyzed can be accessed via \dataset[doi: 10.17909/33qn-6x52]{https://doi.org/10.17909/33qn-6x52}. 

%\acknowledgments

\bibliographystyle{aasjournal}
\bibliography{full}

\end{document}